\documentclass[bibyear]{aa}
\usepackage{graphicx}
\usepackage[varg]{txfonts}
\usepackage{color}
\begin{document}

\title{Interaction of the accretion flows in corona and disk near the
 black hole in AGN.}
% \subtitle{}
\author{E. Meyer-Hofmeister\inst{1} \and B.F. Liu
  \inst{2,3} \and E. Qiao\inst{2,3} }

\offprints{Emmi Meyer-Hofmeister; emm@mpa-garching.mpg.de}

\institute
    {Max-Planck-Institut f\"ur Astrophysik, Karl-
     Schwarzschildstr.~1, D-85740 Garching, Germany
    \and
    Key Laboratory of Space Astronomy and Technology, National
    Astronomical Observatories, Chinese Academy of Sciences, Beijing
    100012, China
    \and
    School of Astronomy and Space Science, University of Chinese
    Academy of Sciences, 19A Yuquan Road, Beijing 100049, China
     }

    \date{Received: / Accepted:}

\abstract 
      {Accretion flows toward black holes can be of a quite different
  nature, described as an optically thick cool gas flow in a
  disk for high accretion rates  or as a hot coronal 
  optically thin gas flow for low accretion rates, possibly affected
  by outflowing gas.
      The detection of broad iron emission lines in active galactic nuclei (AGN) indicates the
  coexistence of corona and disk. The appearance and relative
  strength of such flows essentially depends on their interaction. Liu et al. 
  suggested that condensation of gas from the corona to the disk
  allows to understand accretion flows of comparable strength of
  emission. Matter inflow due to gravitational capture of gas is
  important for the condensation process. We discuss observational
  features predicted by the model.
      Data from simultaneous observations of AGN with
  {\it {Swift's}} X-ray and
  UV-optical telescopes are compared with the theoretical
  predictions.
       The frequent detection of broad iron K$\alpha$ emission lines
  and the dependence of the emitted spectra on the
      Eddington ratio,
  described by the values of the photon index $\Gamma$ and the
  two-point spectral index $\alpha_{\rm{ox}}$ are in approximate
  agreement with the
  predictions of the condensation model; the latter, however, with a large scatter.
  The model further yields a coronal emission concentrated in a narrow inner
  region as is also deduced from the analysis of emissivity profiles. 
       The accretion flows in bright AGN could be
  described by the accretion of stellar wind or interstellar medium
  and its condensation into a thin disk.
    
 }

\titlerunning{Accretion flows in corona and disk in AGN}
\authorrunning{E. Meyer-Hofmeister, B.F.Liu and E.Qiao}

\maketitle

\section{Introduction}

Observations indicate that the accretion of gas onto black holes
can be described in the form of an optically thick, geometrically thin
accretion disk or a hot optically thin, vertically extended 
flow. In the inner regions of active galactic nuclei (AGN) both types of accretion flow often
exist, that is, a disk embedded in a hot coronal flow.   X-ray
contributions from a corona have already been found in
spectra during early AGN observations (e.g., Elvis et al. 1994, Nandra \& Pounds
1994). Relativistic Fe K$\alpha$ lines were first detected in
ASCA data (Tanaka et al. 1995). The presence of X-ray reflection in
the spectra of many AGN indicates that the downward emission from
the hot coronal flow is Compton scattered by matter of an accretion
disk in the inner region (Fabian et al. 2000). Recent results, for example, a
multi-epoch spectral analysis of all {\it XMM-Newton, Suzaku,} and
{\it NuSTAR} observations of Mrk 335 by Keek \&
  Ballantyne (2016),  confirm the existence of the accretion disk
  corona.

 For the hot accretion flow different solutions were investigated. 
The advection-dominated accretion flow (ADAF, Narayan \& Yi 1994, 1995a),
the adiabatic inflow-outflow solution (ADIOS, Blandford \& Begelman 1999), 
the convection-dominated accretion flow (CDAF, Igumenshchev \& Abramowicz
1999), and the luminous hot
accretion flow (LHAF, Yuan 2001) have been studied in detail (for a
recent review see Yuan and Narayan (2014).  
Among these solutions, the ADAF is believed to exist in black hole X-ray binaries (BHXRBs) at low/hard state  and low-luminosity AGN.

The coexistence of the cool and the hot accretion flow raises the
 question of the interaction of such flows. 
A radiatively coupled disk corona model was proposed  for AGN by  Haardt \&
Maraschi (1991). It is found that a significant fraction  of the accretion energy should be released in the corona in order to produce strong X-ray emission with spectral index $\sim 1$ in AGN.  A more detailed investigation on the interaction of disk and corona around a black hole was performed by Meyer et al. (2000), where both energy exchange and gas exchange are taken into account. Such a model predicts  how the geometry of the two-phase accretion flows varies with the accretion rate in the innermost area, thereby leading to spectral variation.

The physical processes involved in  the disk-corona interaction are similar
for stellar-mass black holes and supermassive black holes if the mass supply to the accretion flows is the same. 
However, an essential difference between  AGN and BHXRBs is the environment.
In AGN,  matter can be captured from stellar winds of evolved
stars in the central region of the galaxy or from
interstellar medium, in contrast to the Roche-lobe mass overflow in BHXRBs which results in
mass inflow concentrated to the mid-plane. The inflowing gas in AGN is hot and the density in the corona is
large. This causes condensation of coronal gas
to the underlying disk, which, meanwhile,  produces strong X-ray emissions as observed in AGN  (Liu et al. 2015, Qiao \& Liu 2017).

The aim of our
paper is to check whether simultaneous UV/optical and X-ray
observations of suitable AGN with moderate accretion
rates $L_{\rm{bol}}/L_{\rm{Edd}} \le 0.1$ are in agreement with the
condensation model. In Section 2 we give a brief review
of corona models in the context of AGN. 
In Section 3 we
describe the physics of the energy balance
between corona and disk and show numerical results. In Section 4 we
compare characteristic values of the spectral energy distribution
obtained from observations with model predictions. Our conclusions are presented in Section 5.

\section{Interaction of disk and corona} 

Observations in AGN provides strong evidence supporting  interaction
between the disk and corona in the proximity of central black holes.
The X-ray spectrum, which is believed to be composed of direct coronal
emission,  reflection and broadened iron lines, and the reverberation
analyses (for a review see Uttley et al. 2014) indicate strong
coupling between the corona and the disk. The variability of the relative
strength of these radiation components with flux, as observed in some
AGNs (e.g., Gallo et al. 2015), implies a change of the geometry of the
corona flow with Eddington ratio $L_{\rm bol}/L_{\rm Edd}$.   
With the improving quality of spectra from different observatories,
the X-ray emission from coronae and its interaction with the thin disk in the innermost regions of AGN become
a main  astrophysical goal (e.g., Fabian et al. 2016, Wilkins et al. 2014, Kara et al. 2016,
Wilkins et al. 2016, Keek \& Ballantyne 2016) .

Theoretically, the interaction of disk and corona can lead to mass exchange from
corona to disk or vice versa. Evaporation of matter from the disk
causes the truncation of a disk in its inner region from a certain
radius inward to the innermost circular stable orbit
(ISCO), provided that the mass supply to the disk accretion is low
(Meyer et al. 2000, R\'o\.za\'nska \& Czerny 2000, Mayer \& Pringle
2007). It is the vertical heat conduction that plays a key role, which 
establishes the balance between cool disk and hot corona. This
process yields an important feature for
understanding the state transitions and spectra of LMXBs (Remillard \&
McClintock 2006, Yuan \& Narayan 2014, Belloni and Motta 2016).

In the inner region, radiative cooling  can dominate over conduction if the Eddington-scaled accretion rate in the corona is relatively high, $\dot m\ga 0.01$. 
In this case, the corona gas partially condenses into the disk and a weak
inner disk forms underneath the corona. This occurs in BHXBs only for a  few cases (Miller 2006, Rykoff
et al. 2007, Tomsick et al. 2008, Reis et al. 2010) since the corona gas is supplied by disk evaporation, with which the coronal accretion rate  is lower than  0.02 (Liu et al. 2006, Meyer et al. 2007, Taam et al. 2008). However, in
 AGN,  the  accretion rate supplied from wind or interstellar medium to the corona can be higher; condensation of corona gas to a thin disk is then a common phenomenon in bright AGN.

The heating of the corona was a long-existing problem in
  bright AGN.
To solve this problem Haardt and Maraschi (1991) considered a
``two-phase'' accretion disk model,
which was the first work on interaction between disk and
corona in AGN. A substantial fraction of the gravitational power is
assumed to be dissipated via buoyancy and reconnection of magnetic
fields in a hot corona surrounding a disk with Comptonization of soft
photons from the cold phase as the main cooling mechanism (for details see Liu et al. 2002; 2003).
Nakamura \& Osaki (1993) presented a model along
 this scenario emphasizing the strong coupling between the cold disk and the hot
corona.

Recently the disk-corona interaction for bright AGN was investigated by Liu et
al.(2015) and Qiao \& Liu (2017) taking into account the
accretion of matter from stellar winds or interstellar medium in the AGN environment.
A schematic description of the mass and energy flow in the inner region for either
mass supply from vertically extended hot gas (stellar wind and
interstellar medium material) or Roche-lobe overflow in the mid-plane was shown in Liu et
al. (2015, Fig. 1).  The accretion of strong stellar winds or interstellar medium  provides a mechanism for producing strong X-ray emission as observed in AGN.  The energy comes from coronal accretion and condensation-released thermal energy.  This alleviates the long-existing problem of corona heating, which had been a basic assumption in previous accretion models for bright AGN (e.g., Haardt \& Maraschi 1991, Nakamura \& Osaki 1993).  On the other hand, if the gas supply rate by wind or stellar medium is very low, no condensation occurs. Instead, the disk gas evaporates to the corona, leading to a complete depletion of the thin disk. This picture is consistent with the low-luminosity AGN. 
The detailed processes of the
interaction are summarized in the following.

\section{Computational results for condensation in AGN}

The interaction of corona and disk underneath arises through
cooling of electrons. Due to the
large temperature difference between the hot ADAF and the cool disk, thermal conduction and inverse Compton scattering
result in cooling of electrons until the coupling
of electrons and ions becomes efficient with decreasing height and
decreasing temperature. In the layer
between the coupling interface and the disk surface, electrons and ions
have the same temperature, the density rises and radiation losses
become important. If the radiation losses are larger than the
down-flowing heat flux, matter condenses from the corona into the
disk. If the down-flowing flux cannot be radiated away, matter
evaporates from the disk into the corona. The efficiency of the
radiation losses depends critically on the density in the radiative layer and the pressure in the
corona. Thus, the condensation/evaporation rate  is  a function of the mass-flow rate in the corona (Meyer et
al. 2007, Liu et al. 2007).

For simplicity, the accretion flows are divided into three layers in the vertical direction, that is, the disk, the transition layer and the corona.  The upper, vertically extended corona  is similar to an ADAF and can be described by the self-similar solutions (Narayan \& Yi 1995b),
\begin{eqnarray}\label{para}
\begin{array}{l}
p=1.71\times10^{16}\alpha^{-1}c_{1}^{-1}c_{3}^{1/2}m^{-1}\dot m r^{-5/2} \ \ \ \rm g \ cm^{-1} \ s^{-2}, \\
n_e=2.00\times10^{19}\alpha^{-1}c_1^{-1}c_{3}^{-1/2}m^{-1} \dot m r^{-3/2}\ \ \ \rm cm^{-3},  \\
q^{+}=1.84\times 10^{21}\varepsilon^{'}c_{3}^{1/2}m^{-2}\dot
m r^{-4}\ \ \rm ergs \ cm^{-3} \ s^{-1}, \\
c_s^2 =4.50\times10^{20}c_{3}r^{-1} \ \ \rm cm^{2} \ s^{-2},
%\nonumber
\end{array}
\end{eqnarray}
where $p$, $n_e$, $q^{+}$, and  $c_s$  are  the coronal  pressure,   electron density,   viscous heating rate,  and sound speed, respectively. $\alpha$ is the viscous parameter,  $m$  the black hole mass in unit of  solar mass,  
$\dot m$  the coronal accretion rate in unit of   
Eddington  rate, $r$  the radius in unit of
the Schwarzschild radius, and $c_1,c_3,\varepsilon^{'}$ are functions of advection fraction of accretion energy,$f$. 

The temperatures in ions and electrons, and $f$ are determined by three equations, that is, the state equation, the energy balance between heating and cooling for ions and for electrons, respectively.  As $f$ is implicitly contained in  $c_1,c_3$, and $\varepsilon^{'}$, and the energy equations are non-linear, iterative computations are needed to solve the ADAF equations. 

The main difference between the ADAF and corona originates from the additional cooling term in the energy equation caused by vertical conduction to the underlying disk and the Compton scattering of coronal electrons with disk photons. The heat conduction  plays an important role in determining the corona quantities  and mass exchange rate between the disk and the corona. The flux conducted from the corona to the transition layer  can be approximated as (Spitzer
1962),
\begin{equation}
F_c^{ADAF} =k_{0}T_{\rm e}^{5/2}{dT_{\rm e}\over dz}\approx k_{0}T_{\rm e}^{7/2}/H
,\end{equation}
with    $H$ the corona height, $\kappa_0$ the thermal conductivity constant,  $T_{e}$  the electron temperature which is determined by the energy balance between heating by collisions with ions and cooling by radiation through Bremsstrahlung, synchrotron, Compton scattering, complemented by the state equation, and energy balance between viscous/magnetic heating and collisional cooling for ions.

The energy balance in the transition layer is set by the incoming
conductive  flux, Bremsstrahlung radiation flux , and the enthalpy
flux carried by the mass condensation flow, which yields the condensation rate
$\dot m_z$ per unit area at a given distance from the center (see Liu
et al. 2015)
\begin{equation}\label{cnd-general}
\dot m_z= {{\gamma-1} \over \gamma}\beta {{F_{c}^{ADAF}} \over {\Re
T_{i}/ \mu_{i}}}(\sqrt{C}-1),
\end{equation}
with
\begin{equation}\label{C}
C \equiv\kappa{_0} b \left(\frac{\beta^2 p^2}{\pi k^2}\right)
\left(\frac{T_{\rm {cpl}}}{F_c^{\rm{ADAF}}}\right)^2,
   \end{equation}
where  $T_{i}$ and $ \mu_{i}$ are temperature and molecular
weight of ions, $\beta$ is the ratio of gas pressure to total pressure,
$\gamma$ is the ratio of specific heats $\gamma={8-3\beta\over 6-3\beta}$, 
$T_{\rm cpl}$  is the coupling temperature in the transition layer,
determined by taking the heating of the ions as completely shared with
the electrons,  $b$ is the coefficient of free-free radiation (Sutherland \& Dopita 1993)
and $k$ is the Boltzmann constant.
 
 As the condensed  gas is the only source  for disk accretion, the
 emission from the disk  depends on both condensation rate and
 irradiation from the corona. This disk emission is the main part of
 seed photons in Compton cooling in the corona. Therefore, $p$ and
 $F_c^{ADAF}$ on the right side of Eq.(\ref{cnd-general})  are
 implicit functions of the condensation rate $\dot m_z$  resulting
 from the energy balance.  Numerical computations consistently yield
 the condensation rate, the coronal temperatures and densities for a 
 given  black hole mass, accretion rate $\dot m\equiv\dot M/\dot
 M_{\rm Edd}$, viscous parameter $\alpha$ and magnetic parameter
 $\beta$.

 We then calculate the spectra emitted from both the corona and
 disk. Repeating such calculations for a series of accretion rates,
 we can obtain the variation of the spectrum with the Eddington ratio.

With typical parameters, $M=10^8 M_\odot$, $\alpha=0.3$, $\beta=0.95$,  and $\dot m=0.015-0.1$, we perform numerical computations.
 For the illumination of the disk by the corona we
assume that the radiation of the somewhat extended corona can be
approximated by the radiation from a point
source at a height $10R_S$ (Qiao \& Liu 2017). To check the effect of
the simplification we discuss the distribution of coronal
luminosity which depends on the mass flow. 

The computations show that the hot gas
will partially condense to an underlying inner disk, while a certain amount
remains in the corona. The accretion rate in the disk decreases outward to a radius $r_{\rm{d}}$ where no condensation is
found anymore. As there is no "Roche-lobe overflow" to the disk,  the disk ends at $r_{\rm{d}}$.
For accretion rates around 0.02 the resulting values for $\nu L_{\nu}$
are comparable in the optical/UV and
the X-ray domain, documenting the presence of a hot and a cool flow, a
precondition for broad iron lines. For a rate $\dot
m_{\rm{Edd}}$=0.03 the value in the optical/UV range becomes
higher than that in the X-ray range (see Qiao \& Liu 2017).  An inner
disk should exist only if the accretion rate is not very low. 

\begin{figure}
    \centering
   \includegraphics[width=8.5cm]{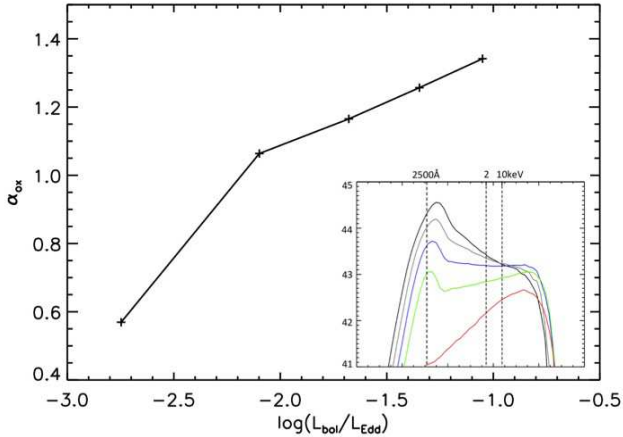} 
   \caption {Optical--X-ray spectral index
     $\alpha_{\rm{ox}}$ as a function of the Eddington ratio for $M=10^8 M_\odot$,
     $\alpha=0.3$, $\beta=0.95$,  "+" values for mass supply rates $\dot m$=
     0.015, 0.02, 0.03, 0.05, 0.1.
   The inset shows spectra, log[($\nu L_\nu(\rm{erg
         s^{-1}})$] plotted
   against log[$\nu(\rm{Hz})]$
   for the same rates $\dot m$ from the bottom up
     }
    \label{f:alpha-ox}  
   \end{figure}

 The two-point spectral index of optical/UV and X-ray radiation
 $\alpha_{\rm{ox}}=-log[L(\nu_{\rm{X-ray}})/L(\nu_{\rm{UV}}]/log(\nu_{\rm{X-ray}}/\nu_{\rm{UV}})$
(Tananbaum et al.1979), here at $2500 A^\circ$ and 2keV respectively, documents the relative  strength of corona to
disk (results for many AGN compiled by Steffen et
al. 2006).  Fig.\ref{f:alpha-ox} shows how the index $\alpha_{\rm{ox}}$
varies with the Eddington ratio $L_{\rm bol}/L_{\rm Edd}$.  
 
 The Figure shows  that $\alpha_{\rm ox}$ increases with the Eddington
 ratio, which means the relative strength of corona emission to disk
 emission is weaker at higher Eddington ratios. For a higher mass supply
 rate to the corona, condensation is more efficient and starts at a
 larger distance. This results in less gas remaining in the inner
 corona and hence a decreased X-ray emission.  Therefore,  the
 relative strength of the corona decreases with increasing Eddington
 ratio. The steep variation of $\alpha_{\rm ox}$  with $L_{\rm
   bol}/L_{\rm Edd}$ at  $\dot m\sim 0.02$  is caused by the
 transition from a pure ADAF to a disk-corona structure.
 
\begin{figure}
    \centering
     \includegraphics[width=8.5cm]{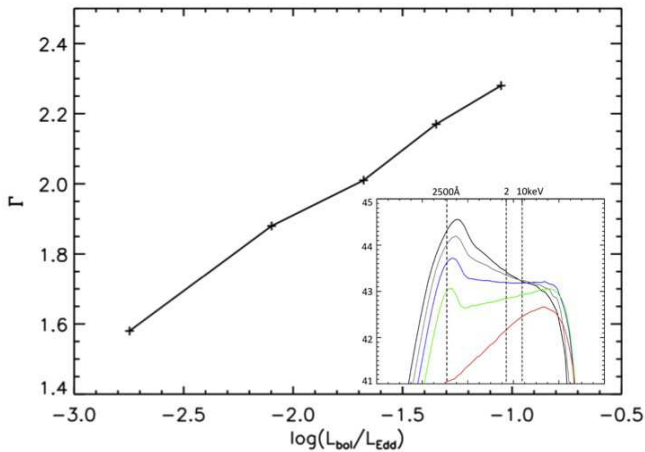}
       \caption{ Hard X-ray (2-10keV) photon index
         $\Gamma$ as a function of the Eddington ratio, 
           parameters and inset as in Fig. \ref{f:alpha-ox}.}
     \label{f:gamma}  
   \end{figure}

We also calculate the hard  X-ray (2-10keV) photon index  for various Eddington ratios  and plot the results in Fig.\ref{f:gamma}.
 The model predicts quite a large range of photon index, from
 1.6 to 2.3, for an Eddington ratio change from 0.002 to 0.1. Very 
 low photon indices as $\Gamma\sim1.6$, correspond to a
 hard-state spectrum. In this case, the accretion flow is dominated by
 the corona and the inner disk is very small and weak. This is because
 cooling in the corona is inefficient at a low mass supply rate, thus,
 condensation can occur only in the region very close to the ISCO, while
 very little gas can condense to the disk.   With increase of mass
 supply to the corona, cooling becomes more
 efficient, leading to more condensation. This puts the main
 accretion flow into the disk instead of the corona while the X-ray
 spectrum becomes steeper.  At an accretion rate $\dot m\sim 0.1$, 
 a large fraction of the coronal gas condenses to the disk. The corona
 becomes weak and a high value of $\Gamma$ is reached.
 
   The total amount of mass flow in the disk at given distance $r$ is
   calculated by integration of the condensation rate from the outer edge
   of the disk at $r_{\rm{d}}$ to $r$. The mass flow remaining in
   the corona then follows from the difference between the rate of
   inflowing matter in the corona and the integrated condensation
   rate. The mass flow in the corona is advection dominated
   (Narayan \& Yi 1995b) and the radiation of the hot flow is due to the
   various electron cooling processes, Bremsstrahlung cooling,
   synchrotron cooling, self-Compton cooling and cooling by
   underlying disk photons on the electrons in the corona.

\begin{figure}
   \centering
   \includegraphics[width=8.5cm]{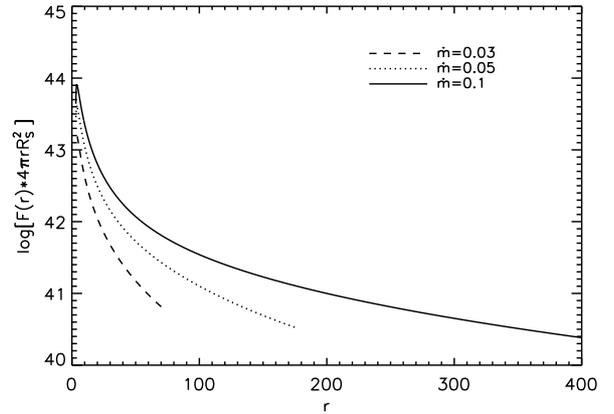}
   \caption{Luminosity emitted from the coronal annulus ($\Delta r=1$)
     along distance for  mass supply rates $\dot m=0.03,0.05,0.1$}
  \label{f:luminosity}  
\end{figure}

The luminosity of the corona is then calculated assuming irradiation of the
disk from a point source at a height of 10$R_{\rm{S}}$ (Schwarzschild
radius). Though the condensation leads to a somewhat radially extended
mass distribution of the corona it was found that most of the
luminosity is released in a very narrow inner region 
and might be simulated by radiation from a point source.
We show the luminosity distribution in Fig.\ref{f:luminosity}.

\begin{figure}
   \centering
   \includegraphics[width=8.5cm]{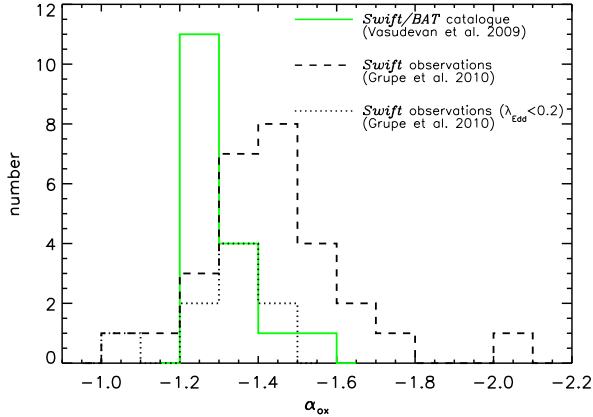}
   \caption{Histogram of values $\alpha_{\rm{ox}}$, relation between
     optical/UV and X-ray emission for two samples, sources with black
     hole mass $10^{7.5} \le M_{\rm{BH}}/M_\odot \le 10^{8.5}$.
   (a) sample of AGN (Vasudevan et al. 2009), with $\lambda_{\rm{Edd}}
     \le 0.1$, solid line,
     (b) sample of AGN (Grupe et al. 2010), bright soft X-ray
     selected AGN, dashed line,
     and sample subset of sources with $\lambda_{\rm{Edd}} \le 0.2$,
     dotted line.}
   \label{f:hist}
   \end{figure}

{\section{Comparison with observations}

\subsection{Optical/UV to X-ray luminosity in active galactic nuclei}

The index $\alpha_{\rm{ox}}$ has a potential power for the
documentation of accretion flows.
As the dependence of flux at $2500 A^\circ$  on the black hole mass  differs from that  at X-ray,  the value of $\alpha_{ox}$  can be different for objects with different black hole mass even if their Eddington ratios are the same.  This may lead to scattering when the model predictions are compared with observations for samples with a large range of black hole masses. 
Of special interest for a comparison
of theoretical results and observations are simultaneous observations 
of optical/UV and X-ray flux such as those by Vasudevan et
al. (2009) for 26 low-absorption AGN from the {{\it{Swift}-BAT}}
9-month catalog with $\lambda_{\rm Edd}=L_{\rm{bol}}/L_{\rm{Edd}} \le
0.1$ which is the parameter range studied in the condensation model.
For the comparison with model results ($M_{\rm{BH}}/M_\odot =10^8$) we
consider sources with black hole masses in the range
$10^{7.5}\le M_{\rm{BH}}/M_\odot \le 10^{8.5}$ (17 sources).
A histogram of these values is shown in
Fig.\ref{f:hist}. Most of the sources lie in the range 1.1 to 1.35 as
predicted by the model (Fig.\ref{f:alpha-ox}). From the model, a
correlation between $\alpha_{\rm{ox}}$ and the Eddington ratio
would be expected, but was not found in the studied
sample, maybe due to inaccuracies
in mass determination methods as Vasudevan et al. (2009) pointed out.

A second sample with simultaneous observations are bright soft
X-ray-selected AGN (Grupe et al. 2010), where 38
sources have a black hole mass around $10^8M_\odot $. Values of $L_{\rm{
  bol}}/L_{\rm{Edd}}$ derived from the SED lie in a wide range; from 9 sources having $L_{\rm  bol}/L_{\rm{Edd}} \le 0.2$ leading up to sources having as high as 10  $L_{\rm  bol}/L_{\rm{Edd}}$ . The
full sample and the subset are shown in Fig.\ref{f:hist}. 
The difference between the sample and the subset might indicate an increase of $\alpha_{\rm{ox}}$ with increasing  Eddington ratio. This is in agreement with the model prediction.

A third sample are X-ray selected type 1 AGN from the
{\it{XMM-COSMOS}} survey studied by Lusso et al. (2010).
The relation between $\alpha_{\rm{ox}}$ and $\lambda_{\rm{Edd}}$ is
shown in their Fig. 12. About half of the sources lie in the range of
$\lambda_{\rm{Edd}}$ 
0.01 to 0.1, as considered in the condensation model. There is a
large scatter of $\alpha_{\rm{ox}}$ values from 1.15 to 1.7. The
scatter could be due to inaccuracies in mass determination and the
determination of the SED. A comparable scatter is also found for the
dependence of the hard bolometric correction on $\lambda_{\rm{Edd}}$ (their Fig. 11).

 A very recent study of this relation (Fig.7 in Liu H. et
  al. 2017) also indicates no obvious correlation for many different
  samples.  However,  a correlation between $\alpha_{ox}$ and black
  hole mass has been found in their samples. This implies that the
  correlation between $\alpha_{ox}$ and Eddington ratio could
  become blurred  by the different black hole masses in   the above investigations. The large uncertainty in determining the X-ray flux (by hardness ratio rather than X-ray spectrum) is another reason.

But the study of the relation between $\alpha_{\rm{ox}}$
and the monochromatic luminosity $L_{2500 A^\circ}$ gives clearer results.}
Lusso et al. (2010) show this relation
for the sources of their sample and find a significant correlation,  
though with quite a scatter also (their Fig. 7). The derived best-fit relation
is 
\begin{equation}
\alpha_{\rm{ox}}(L_{2500{\AA}})=(0.154\pm 0.010)log(L_{2500 {\AA}}-
(3.176 \pm 0.223)  
.\end{equation}

The theoretical calculated values are close to this relation: for
the mass supply rate of, for example, 0.02 or 0.1, the values of log
[$L_{2500{\AA}}] [\rm{erg s^{-1} Hz^{-1}}]$
are 27.9 and 29.2 and values of
$\alpha_{\rm{ox}}$ are 1.06 and 1.34. The best-fit relation gives
$\alpha_{\rm{ox}}(L_{2500{\AA}})$=1.12 and 1.34. This means the
theoretical values agree with the best-fit values.

\subsection{The photon index $\Gamma$} 
The hard X-ray photon index of a power-law spectrum in the 2-10 keV band is one
of the main characteristic features of the description of spectral
properties. For a large part of the
sources in the Vasudevan et al. (2009) sample (with low accretion ratio
 as mentioned above) the photon index derived is between 1.8 and
2.0. The condensation model  (Qiao \& Liu 2017) predicts an increase
of the photon index $\Gamma$ from 1.6 to 2.3 for increasing 
Eddington ratio from 0.002 to 0.1 (Fig.\ref{f:gamma}).
The inaccuracies in mass-determination methods make a detailed study of the
dependence of the photon index on the Eddington ratio difficult; for example, for
five sources in two samples: in the Vasudevan et al. (2009) and also the Lubi\'nski
et al. (2016) sample, the black hole masses used differ by more than a
factor of two, which corresponds to an essential difference of values for
$\Gamma$.

The Seyfert 1 galaxy Mrk 335 was observed several times during
recent years. For the low-flux epoch in 2013 ({\it{Suzaku}}
observation) a hard photon index of 1.9 was derived by Wilkins \&
Gallo (2015), which seems in agreement with model
predictions.  Keek \& Ballantyne (2016) took observations of Mrk 335 during
the years 2000 to 2014 and analyzed 12 X-ray
spectra that span nearly a factor of 10 in flux.
The authors point out that the derived values of photon
index strongly correlate with the Eddington ratio (their Fig. 4). The
Eddington ratio was calculated using a bolometric correction depending
on $\Gamma$.

For the
low flux states, they found very low hard photon indices of around
1.4, definitely lower than those derived by Wilkins \& Gallo
(2015)  and predicted by the model.
Theoretically it is difficult to
understand the very low values around 1.4.
Though the photon index decreases with decreasing Eddington ratio as
shown in Fig.\ref{f:gamma},  this trend is inverted for further
decreasing Eddington ratios. This is because condensation no longer occurs
at very low accretion rate. Instead, evaporation depletes the inner
disk, leaving only a corona/ADAF. Thus, the photon index
emitted by an ADAF would increase with decreasing accretion rate, never
reaching a value of $\sim$ 1.4.
From the disk corona interaction model a lower
photon index would be derived for a stronger heating of the corona,
for example, if one assumes a larger viscosity value, 
which  results in less efficient condensation and  a denser corona,
which then emits an X-ray spectrum with smaller photon
index.

\subsection{Illumination of the disk by the corona}
 The pattern of illumination of the disk by an X-ray-emitting
 source is one of the most interesting features of AGN. The
 emissivity profile is composed by the different profiles of the broad
 K$\alpha$ emission line in the reflection spectrum
when emitted from successive radii (Wilkins \& Fabian 2011).
Theoretical emissivity profiles were derived for illumination by
coronae with different extension. The comparison of
theoretical and observed emissivity profiles (Wilkins \&
Fabian 2012) yields, in combination with X-ray reverberation lags,
information on the geometry of the accretion flow.
   
The analysis of theoretical emissivity profiles shows that a
concentration of coronal flux near the center provides a relatively steep
profile in the inner regions (Wilkins \& Fabian 2012).
Using these results Gallo et al. (2015) and Wilkins et al. (2015) investigate 
{\it{Suzaku}} observations for Mrk 335 during the low-flux state in 2013
and conclude from the emissivity profile that the corona is compact,
at low height above the disk and extends to about five gravitational radii.

To compare with predictions of the condensation model one might ask
which kind of emissivity profile would be expected. We show
 the luminosity contributed by radial annuli with $\Delta r=1$ in 
Fig.\ref{f:luminosity}. Most luminosity is released in the inner
region, which should result in a steep inner emissivity profile as
indicated by the observations for Mrk 335 in the low-luminosity state. 
But a detailed determination of the emissivity profile
  is beyond the scope of this paper.

Similar to these results for Mrk 335, Wilkins \& Fabian (2011) point out
that for the narrow line Seyfert 1 galaxy 1H0707-495 the emissivity profile derived for the observation in 2008
(Fabian et al. 2009) suggests a high percentage of the X-ray
flux reflected from the innermost region (Fig. 8).
But it is difficult to estimate the accretion rate in
1H0707-495 at the time of observation. For 1H0707-495 periods of
varying flux are documented (Wilkins et al. 2014).

\subsection{Fe K$\alpha$ lines in AGN}

A frequent detection of relativistic X-ray lines confirms a strong 
interaction between corona and disk flow as predicted by the
condensation model. The broadening of the line indicates that the
reflection occurs in the innermost region.
 
\begin{figure}
   \centering
   \includegraphics[width=8.5cm]{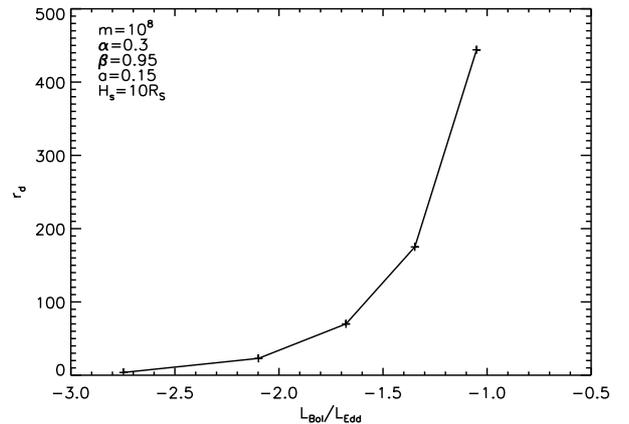}
   \caption{The radial extension of the thin disk for
     mass supply rates $\dot m$ 0.015 to 0.1, shown as a function of Eddington ratio.}
   \label{f:Rout}
 \end{figure}

 For the geometry of disk plus corona, the condensation model
   predicts a 
 coexistence of disk and corona for mass supply rates larger than 0.01
 for the given parameters in this work. For a higher mass supply rate,
 the accretion rate via the thin disk and the disk extension increase
 with the mass supply rate.  In Fig.\ref{f:Rout} we show the disk size
 as a function of Eddington ratio.   
As shown,  the disk extension can increase from a few to several
hundred Schwarzschild radii. For an even higher Eddington ratio
(beyond our calculation) a larger thin disk is expected. This
coexistence provides
the necessary condition for the formation of broad Fe lines. The strength and emissivity profile depend on ionization of iron.    

Fabian et al. (2000) have already pointed out that the broadened iron line in the
X-ray band at 6.4 to 6.9 keV is seen in the X-ray spectrum of many AGN
and, in particular, Seyfert galaxies. Nandra et al. (2007) and Miller (2007)
frequently found iron emission lines in different samples. But the
geometry might be different from that in LMXBs in the intermediate hard
spectral state, where the broad iron lines were found during outburst decline (Meyer-Hofmeister \& Meyer 2011).

    Recently Mantovani et al. (2016) critically examined the evidence for
relativistic Fe K$\alpha$ lines in Seyfert galaxies observed with
{\it{Suzaku}}. The results show that the relativistic line is detected
at $>95$\% confidence in all sources observed with high
signal-to-noise ratio, also in sources where such lines were not
found earlier in {\it{XMM-Newton}} observations. The authors conclude that these
lines are a ubiquitous feature in the spectra of Seyfert
galaxies, but are often difficult to detect without very-high-quality
data. These observations point to the coexistence of disk and corona as
predicted by the condensation model for moderate accretion rates.

A further recent study of  the broad Fe K$\alpha$  line  for a large
sample of AGN with well measured optical parameters indicates that the
detection of the broad Fe K$\alpha$  line may strongly
depend on the Eddington ratio (Liu Z. et al. 2016), interpreted as possibly caused by the disk size varying with the Eddington ratio in the condensation model, as shown in Fig.\ref{f:Rout}.  With the disappearance of the thin disk at low Eddington ratio, no broad Fe K$\alpha$ line is expected.

\section{Conclusions}
               
We study the interaction of disk and coronal accretion flows in AGN
that are fueled by stellar winds or interstellar medium.
Different from Roche-lobe overflow in BHXBs, wind accretion in AGN can
support a strong corona at high accretion rates, providing a physical
scenario for interpreting strong X-ray emission observed in bright
AGN.

Numerical calculations of the interacting mass flows in
disk and corona allow us to derive the emitted spectra and reveal that
the optical--X-ray spectral index $\alpha_{\rm{ox}}$ increases with
the Eddington ratio $L_{\rm{bol}}/L_{\rm{Edd}}$,  which implies that
the emission from the corona relative to the disk is weaker at a
higher Eddington ratio.

We compare the computational results with observations.
 The values of  $\alpha_{\rm{ox}}$ for different samples plotted
 against the Eddington ratio have a large scatter which might be due
 to different black hole mass and inaccuracies of  the SED
 , but theoretical results for a relation between
 $\alpha_{\rm{ox}}$ and luminosity at 2500${\AA}$ are, despite also
 some scatter, in good agreement with best-fit parameters
 (Lusso et al. 2010). 
The dependence of $\Gamma$ on the
theoretically derived Eddington ratio is in agreement with results by Keek \& Ballantyne (2016) for Mrk 335.

The steep inner part of the emissivity profile of Mrk 335 in the
low-luminosity state was interpreted as pointing to illumination by a centrally
concentrated light source (Gallo et al. 2015). A similar illumination
of the disk is predicted by the condensation model with a coronal luminosity released mainly in an innermost region.

A further interesting feature is the apparently frequent detection of
broad Fe K$\alpha$ lines in AGN for a wide range of Eddington ratios. The condensation model with wind accretion provides a natural
interpretation for the occurrence of the relativistic iron lines.

\begin{acknowledgements}
B.F.L. acknowledges the support from  the National Program on Key Research and Development Project (Grant No. 2016YFA0400804) and the National Natural Science Foundation of China (grant 11673026). 
\end{acknowledgements}

{}

\end{document}